\documentclass[3p,times,procedia]{elsarticle}
\usepackage{nupha_ecrc}

\volume{00}

\firstpage{1}

\journalname{Nuclear Physics A}

\runauth{Ramona Lea (for the ALICE Collaboration)}

\jid{nupha}

\jnltitlelogo{Nuclear Physics A}

\usepackage{amssymb}
\usepackage{lineno}

\usepackage[figuresright]{rotating}

\newcommand{\dbar}   {$\overline{\mathrm{d}}$}

\newcommand{\PbPb}   {Pb--Pb}

\newcommand{\mom}    {\mbox{\rm MeV$\kern-0.15em /\kern-0.12em c$}}
\newcommand{\gmom}   {\mbox{\rm GeV$\kern-0.15em /\kern-0.12em c$}}
\newcommand{\mass}   {\mbox{\rm GeV$\kern-0.15em /\kern-0.12em c^2$}}
\newcommand{\Mmass}  {\mbox{\rm MeV$\kern-0.15em /\kern-0.12em c^2$}}
\newcommand{\pt}     {$p_{\rm T}$}
\newcommand{\s}      {$\sqrt{s_{\mathrm{NN}}}$}

\newcommand{\vtwo}   {$v_{2}$}

\begin{document}

\begin{frontmatter}

\dochead{}

\title{(Anti-)deuteron production and anisotropic flow measured with ALICE at the LHC}

\author{Ramona Lea (for the ALICE Collaboration)}

\address{Dipartimento Di Fisica, Universit\`a di Trieste e INFN, Sezione Trieste}

\begin{abstract}
The high abundance of (anti-)deuterons in the statistics gathered in Run 1 of the LHC and the 
excellent performance of the ALICE setup allow for the simultaneous measurement of the elliptic 
flow and the deuteron production rates with a large transverse momentum (\pt) reach.
The (anti-) deuterons are identified using the specific energy loss in the time projection 
chamber and the velocity information in the time-of-flight detector. 
The elliptic flow of (anti-)deuterons can provide insight into the production mechanisms of 
particles in heavy-ion collisions. Quark coalescence is one of the approaches to describe the 
elliptic flow of hadrons, while the production of light nuclei can be also depicted as a 
coalescence of nucleons. In these proceedings, the measured \vtwo\ 
of deuterons produced in \PbPb\ collisions at \s~=~2.76~TeV will be compared to expectations from 
coalescence and hydrodynamic models.
\end{abstract}


\end{frontmatter}


\section{Introduction}
\label{sec:introd}
Anisotropic flow \cite{Ollitrault:1992bk} studies can probe the nature of matter
produced in heavy-ion collisions. These studies allow for the investigation of 
collective effects among produced particles. The angular distribution of all 
the reconstructed charged particles can be expanded into a Fourier series w.r.t.
symmetry plane $\Psi_{\{n\}}$:

\begin{equation}
E \frac{\rm d^3 N}{\rm {d^3\textbf{p}}} = \frac{1}{2\pi} \frac{\rm d^2 N}{p_\mathrm{T} \mathrm{d} p_{\rm{T}} \mathrm{d}y} \left( 1 + \sum_{n=1}^{\infty} 2 v_n \cos \left[ n \left( \varphi -  \Psi_{\{n\}} \right) \right] \right)
\end{equation}
where $E$ is the energy of the particle, \textbf{p} the momentum, \pt\ the transverse momentum, $\varphi$ 
the azimuthal angle, $y$ the rapidity, $\Psi_{\{n\}}$ the symmetry plane angle and  
\begin{equation}
v_n = \langle \cos \left( n(\varphi - \Psi_{\{n\}}) \right)\rangle .
\end{equation}
The second term of the Fourier series (\vtwo) is called elliptic flow and is a 
parameter which may provide insight on initial conditions and particle production mechanisms. For identified 
hadrons \vtwo\ is sensitive to the partonic degrees of freedom
in the early stages of the evolution of a heavy-ion collision
\cite{Voloshin:1994mz}. 
The deuteron is a composite p+n bound state, whose binding 
energy ($\sim$~2.24~MeV) is much lower than the hadronization temperature. 
Thus, it is likely that it would suffer from medium induced breakup in the 
hadronic phase, even if it was produced at hadronization. 
The \vtwo\ measurements for d and \dbar\ provide an important test for the 
universal scaling of elliptic flow~\cite{Nonaka:2003ew} since its \vtwo\  
should be additive both with respect to the \vtwo\ of its constituent 
hadrons and with respect to the \vtwo\ of the constituent quarks of 
these hadrons, i.e. $n_q = 2\times 3$. \\
In these proceedings, the measurements of (anti-)deuterons \vtwo\ are 
compared to other identified particles~\cite{Abelev:2014pua} and to a 
model based on hydrodynamics~\cite{Huovinen:2001cy, Adler:2001nb}.  
\section{Analysis Details}
\label{sec:datasample}
\begin{figure}[!htbp]
\begin{tabular}{ccc}
\begin{minipage}{.33\textwidth}
\centerline{\includegraphics[width=1\textwidth]{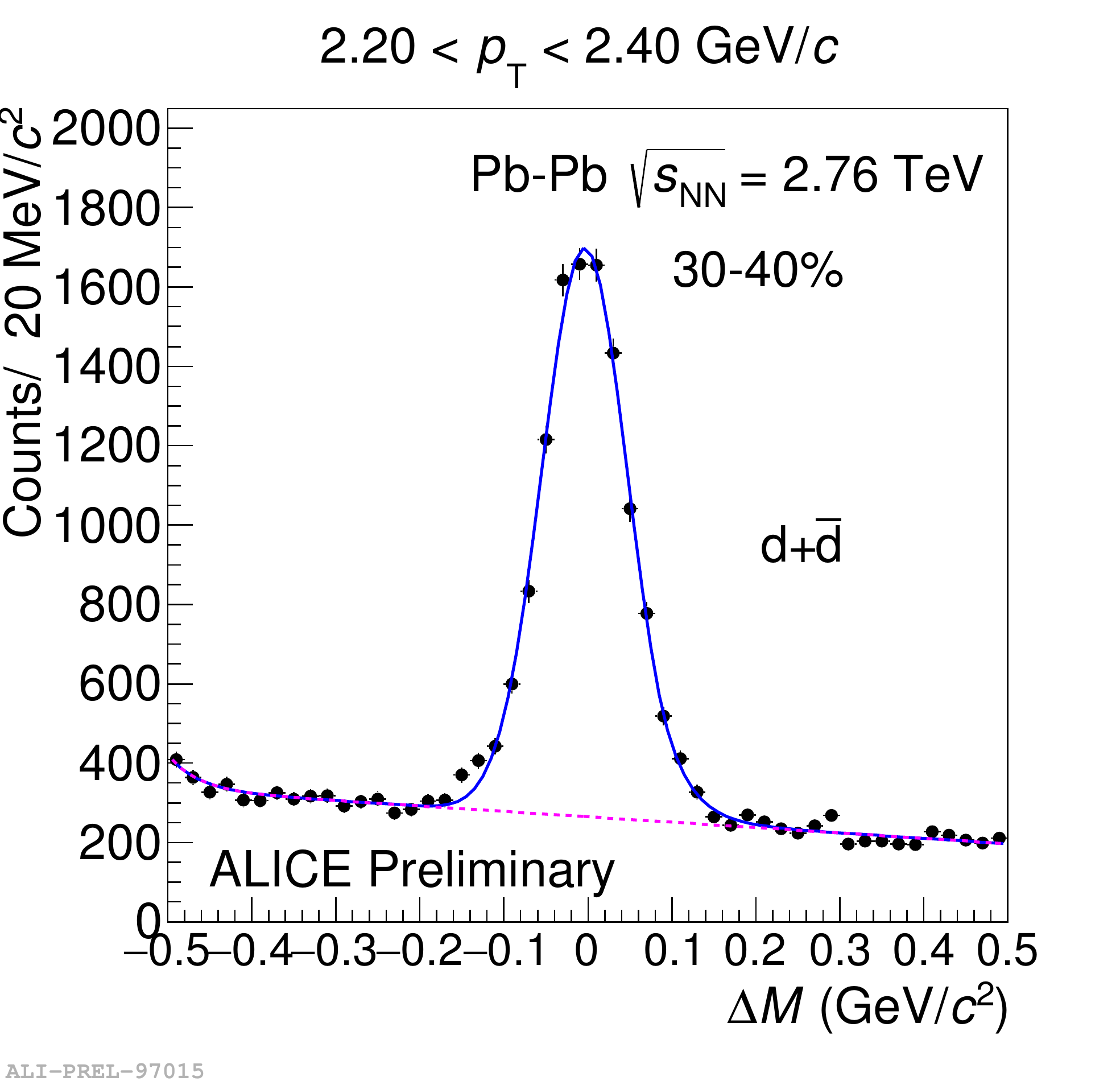}}
\end{minipage} & 
\begin{minipage}{.33\textwidth}
\centerline{\includegraphics[width=1\textwidth]{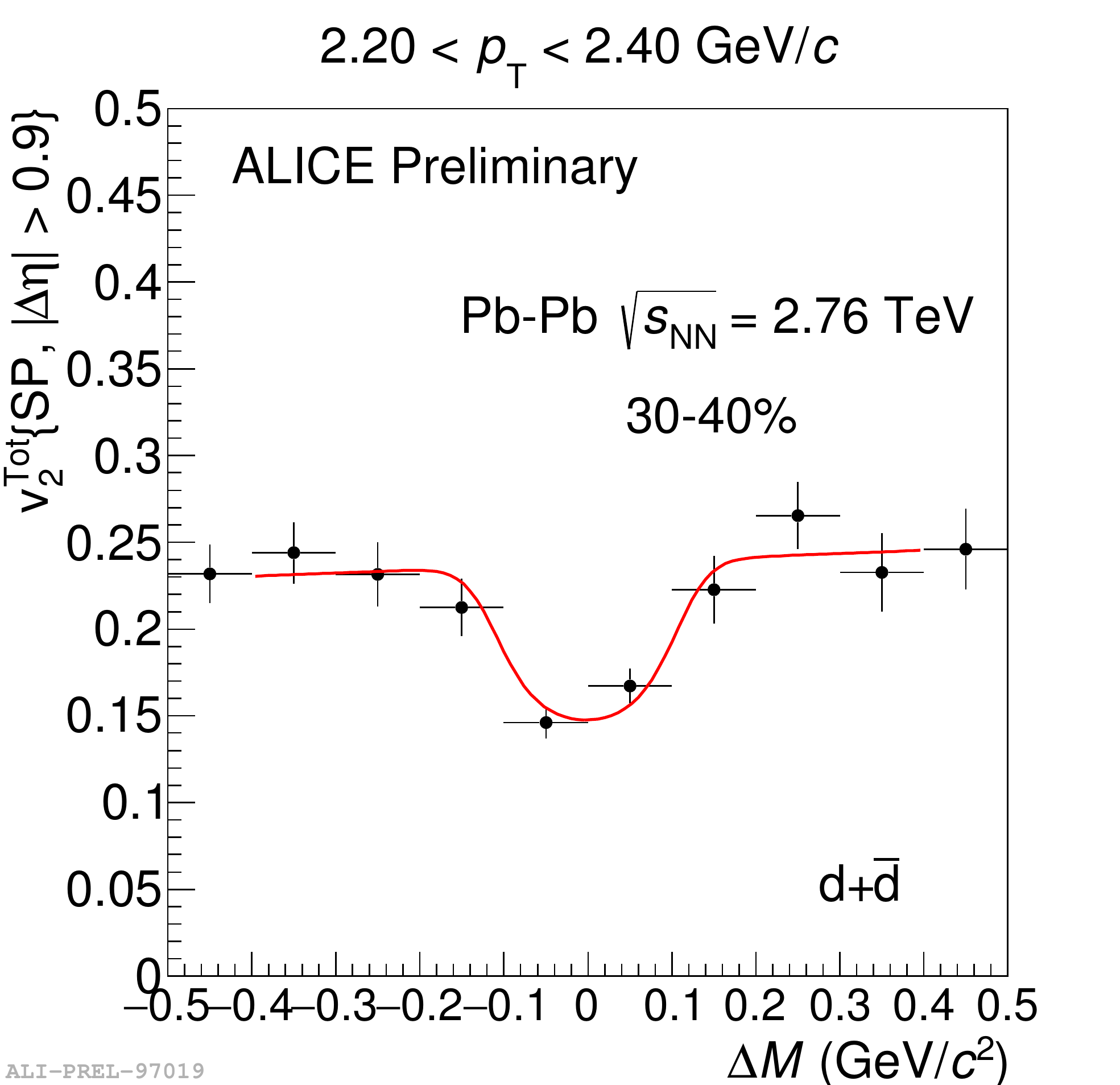}}
\end{minipage} & 
\begin{minipage}{.33\textwidth}
\centerline{\includegraphics[width=1\textwidth]{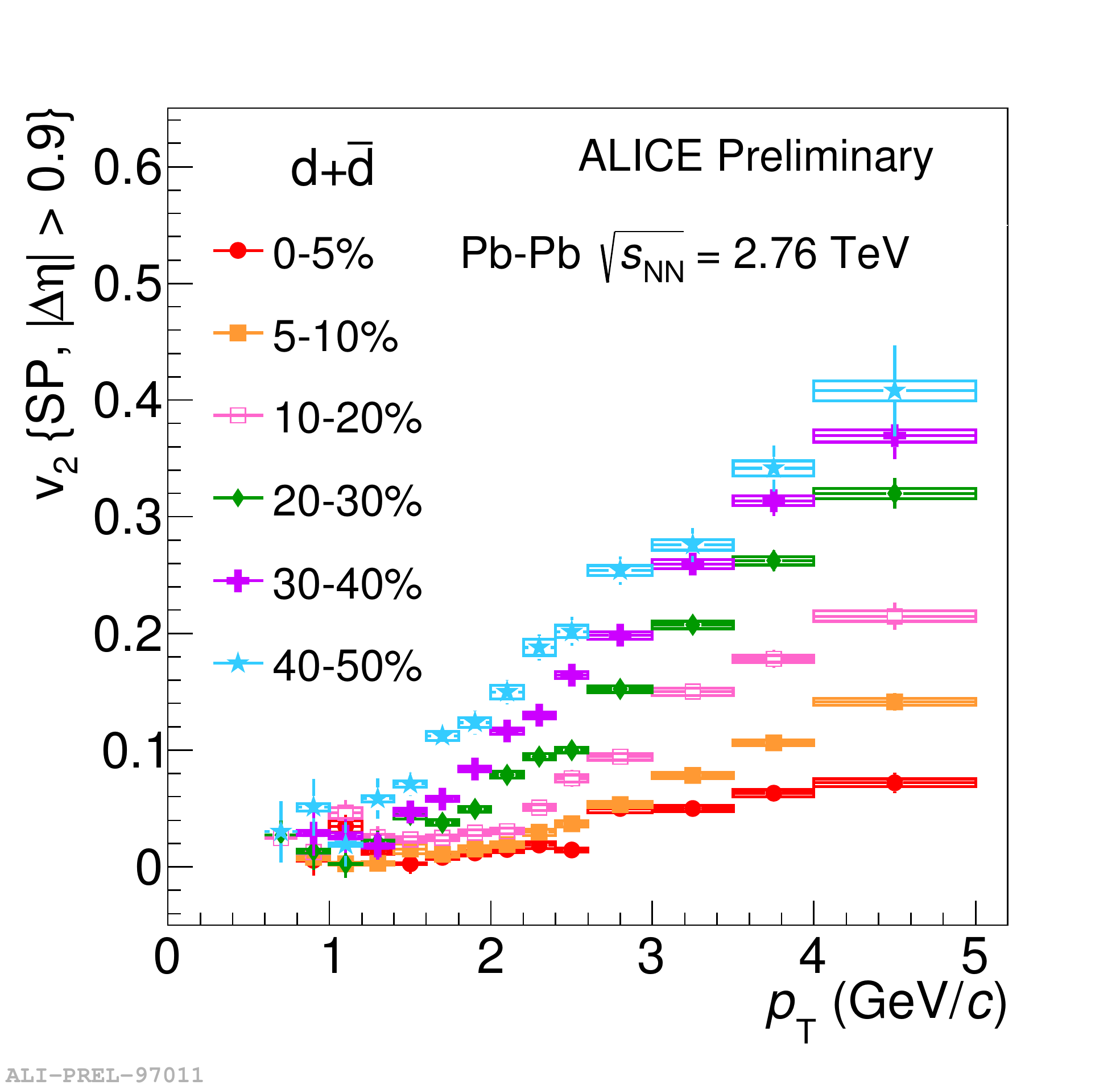}}
\end{minipage} 
\end{tabular}
\caption{Left: Distribution of $\Delta \rm{M}$ for (d+\dbar) in the $2.20 <$~\pt~$<2.40$~\gmom\  
and centrality interval 30-40\% fitted with a function which is the sum of Gaussian plus and exponential 
used to reproduce the signal and an exponential to reproduce the background. 
Middle: The \vtwo$^{\rm{Tot}}$ vs $\Delta \rm{M}$ for (d+\dbar) for 2.20 $<$\pt$<$ 2.40 \gmom\ 
in events with 30-40\% centrality. Points represent the measured $v_2^{\mathrm{Tot}}$, 
while the curve is the fit performed using equation~\ref{eq:v2tot}. 
Right: Measured \vtwo\ as a function of \pt\ for (d+\dbar) in \mbox{\PbPb} collisions 
at \s~2.76~TeV. Lines represent statistical errors, while boxes are systematic uncertainties.
Figure copyright CERN, reproduced with permission.}
\label{fig:method}
\end{figure}

For the analyses presented here, a sample of about 35 million
\PbPb\ collisions at \s\ = 2.76 TeV collected by ALICE \cite{Abelev:2014ffa} in 2011 were used. 
The events were classified into 6 different centrality intervals, which were 
determined using the forward V0 \cite{Abbas:2013taa} scintillator arrays. 
Particle tracking is done by means of the Time Projection Chamber (TPC) \cite{Alme:2010ke} 
and the Inner Tracking System (ITS) \cite{Aamodt:2010aa} with full azimuthal coverage 
for $|\eta|<$0.8. The identification of deuteron (d) and anti-deuteron (\dbar) was performed 
in the same way as done to extract deuteron spectra in \PbPb~\cite{Adam:2015vda}.
Specifically, for momenta up to 1 \gmom\ the energy loss in the TPC gives a clean 
sample of (anti-)deuterons by requiring a maximum deviation of the specific 
energy loss of 3$\sigma$ with respect to the expected signal. Above 1 \gmom\ a Time of Fight (TOF) 
\cite{Abelev:2014ffa} hit is required. 
The signal in the TOF detector is fitted with a function which is the sum of a Gaussian with an 
exponential tail, while the background is fitted with an exponential. 
As an example of the $\Delta \rm{M}$, 
where \mbox{$\Delta \rm{M}$ = $m_{\mathrm{TOF}}$-$m_{\mathrm{d}_{\mathrm{PDG}}}$}, 
for deuterons and anti-deuterons with $2.20 <$~\pt~$<2.40$~\gmom\ and centrality interval 30-40\% is 
shown in the left part of Figure~\ref{fig:method}.\\ 
The \vtwo\ was measured using the Scalar Product (SP) method \cite{Voloshin:2008dg}. 
The contribution to the measured elliptic flow (\vtwo$^{\mathrm{Tot}}$) due to misidentified
deuterons (\vtwo$^{\rm{Bkg}}$) was removed by studying the azimuthal correlations versus 
$\Delta \rm{M}$. This method is based on the observation 
that, since \vtwo\ is additive, candidate \vtwo$^{\rm{Tot}}$ can 
be expressed as a sum of signal (\vtwo$^{\rm{Sig}}$($\Delta$M)) and background 
(\vtwo$^{\rm{Bkg}}$($\Delta$M)) weighted by their relative yields
\begin{equation}
v_2^{\mathrm{Tot}} (\Delta\rm{M}) = v_2^{Sig}(\Delta\rm{M}) \frac{N^{Sig}}{N^{Tot}}(\Delta\rm{M}) + v_2^{Bkg} (\Delta\rm{M}) \frac{N^{Bkg}}{N^{Tot}}(\Delta\rm{M}),
\label{eq:v2tot}
\end{equation}
where N$^{\rm{Tot}}$ is the total number of candidates, N$^{\rm{Bkg}}$ and 
N$^{\rm{Sig}}$ = N$^{\rm{Tot}}$ - N$^{\rm{Bkg}}$ 
are the numbers of signal and background for a given mass and \pt\ interval.  
The yields N$^{\rm{Sig}}$ and N$^{\rm{Bkg}}$ are extracted from fits to the 
$\Delta \rm{M}$ distributions obtained with the TOF detector for each centrality 
and \pt\ interval. The  \vtwo$^{\rm{Tot}}$ vs $\Delta \rm{M}$ for 
(d+\dbar) for 2.20~$<$\pt$<$~2.40 \gmom\ in events with 
30-40\% centrality is shown in the middle panel of Figure~\ref{fig:method}. 
Points represent the measured $v_2^{\mathrm{Tot}}$, while the curve is the fit performed using 
equation~\ref{eq:v2tot}. The $v_2^{\mathrm{Bkg}}$ was parametrized as a first degree 
polynomial ($v_2^{\mathrm{Bkg}} (\Delta\rm{M}) = p_0 + p_1 \cdot (\Delta\rm{M})$).
The \pt\ range exploited in this analysis is 0.5~$<$\pt$<$~5~\gmom.
The measured \vtwo\ as a function of \pt\ for (d+\dbar) is shown in the right 
panel of Figure~\ref{fig:method}. Figure copyright CERN, reproduced with permission.
The value of \vtwo(\pt) increases progressively from central to semi-central collisions. 
 
\subsection{Comparison with other identified particles}
\begin{figure}[!htbp]
\begin{center}
\includegraphics[width=1\textwidth]{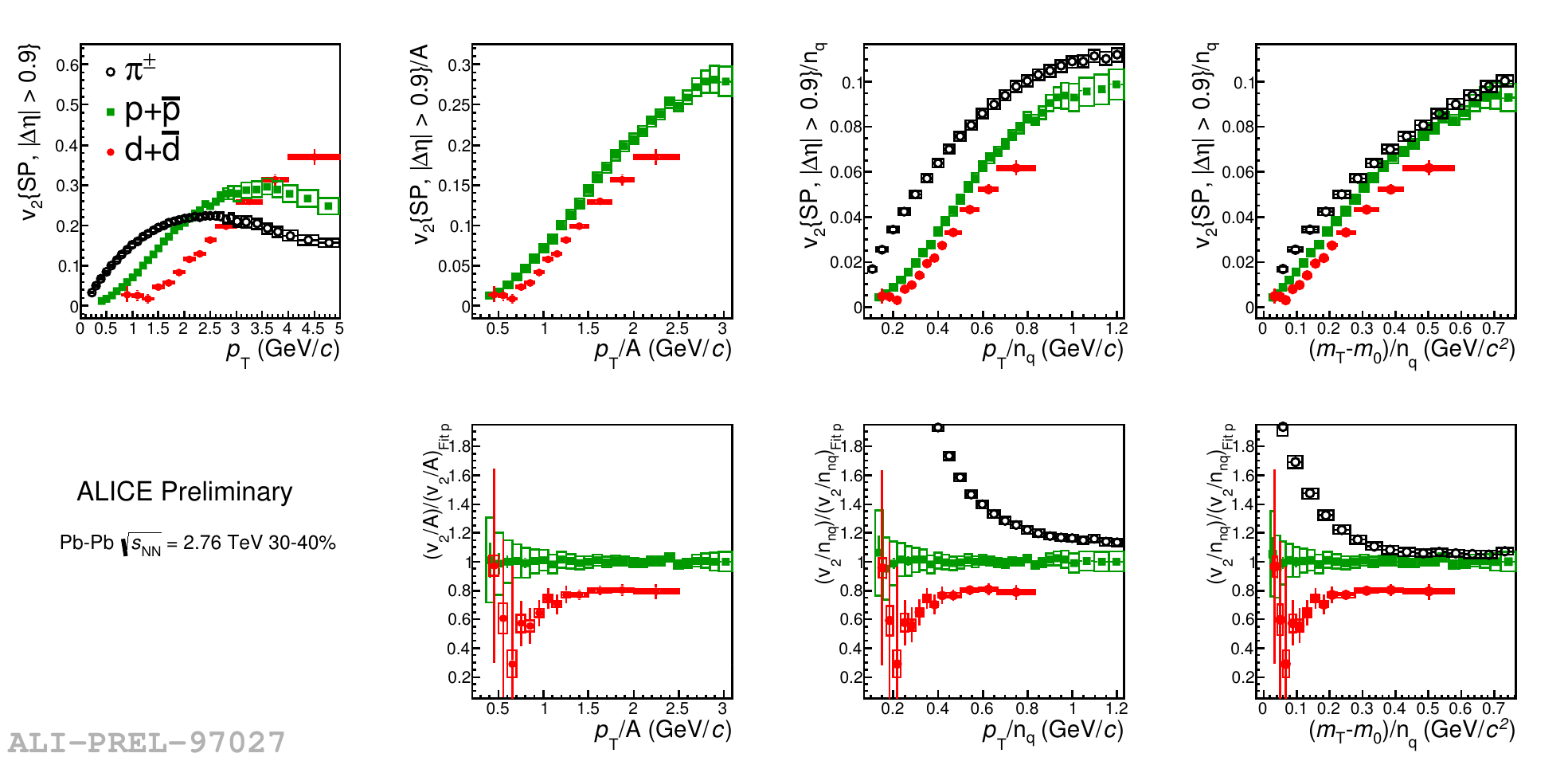}
\caption{Measured (d+\dbar) \vtwo\ compared with the \vtwo\ of other identified 
particles~\cite{Abelev:2014pua} for events with centrality in 30-40\% interval. 
A description of each panel can be found in the text. Figure copyright CERN, reproduced with permission.}
\label{fig:ComparisonMeasured}
\end{center}
\end{figure}

The measured (d+\dbar) \vtwo\ was compared with the \vtwo\ of other identified 
particles~\cite{Abelev:2014pua}. Results for events with centrality in 30-40\% 
are shown in Figure~\ref{fig:ComparisonMeasured}.  
In the top left panel the \vtwo\ vs \pt\ of $\pi^{\pm}$ (empty circles), (p+$\overline{\rm p}$) 
(filled square) and (d+\dbar) (filled circles) as a function of \pt\ are shown.  
It is observed that at low \pt\ deuterons follow the 
mass ordering observed for lighter particles, which is attributed to the interplay 
between elliptic and radial flow. The second panel in the top part of the figure shows  
the \vtwo/A vs \pt/A of (p+$\overline{\rm p}$) (filled squares) and (d+\dbar) (filled circles), 
 while in the second panel of the lower part the ratios between the measured data and a 7$^{th}$ 
  degree polynomial used to describe proton \vtwo\ are reported. For \pt/A$>$1~\gmom\ a deviation from the \pt/A scaling of the order of 20\% is observed. 
  A similar deviation is observed for all the measured centrality intervals. 
The third panels (upper and lower part) are used to test the n$_{\rm q}$ (number of constituent quark) scaling: 
both \pt\ and measured \vtwo\ were scaled by n$_{\rm q}$. Also in this case a deviation 
for the n$_{\rm q}$ scaling of the order of $\approx$20\% 
for each centrality interval is observed. Finally, in the upper fourth panel, the measured \vtwo/n$_{\rm q}$ 
is shown as a function of $EK_T/\mathrm{n}_{\rm q} = (m_{T} - m_0)/\mathrm{n}_{\rm q}$ of each particle:  
significant deviations from n$_{\rm q}$ scaling are seen in data for deuterons. 

\subsection{Comparison with Blast-Wave model}
\begin{figure}[!htbp]
\begin{center}
\includegraphics[width=0.75\textwidth]{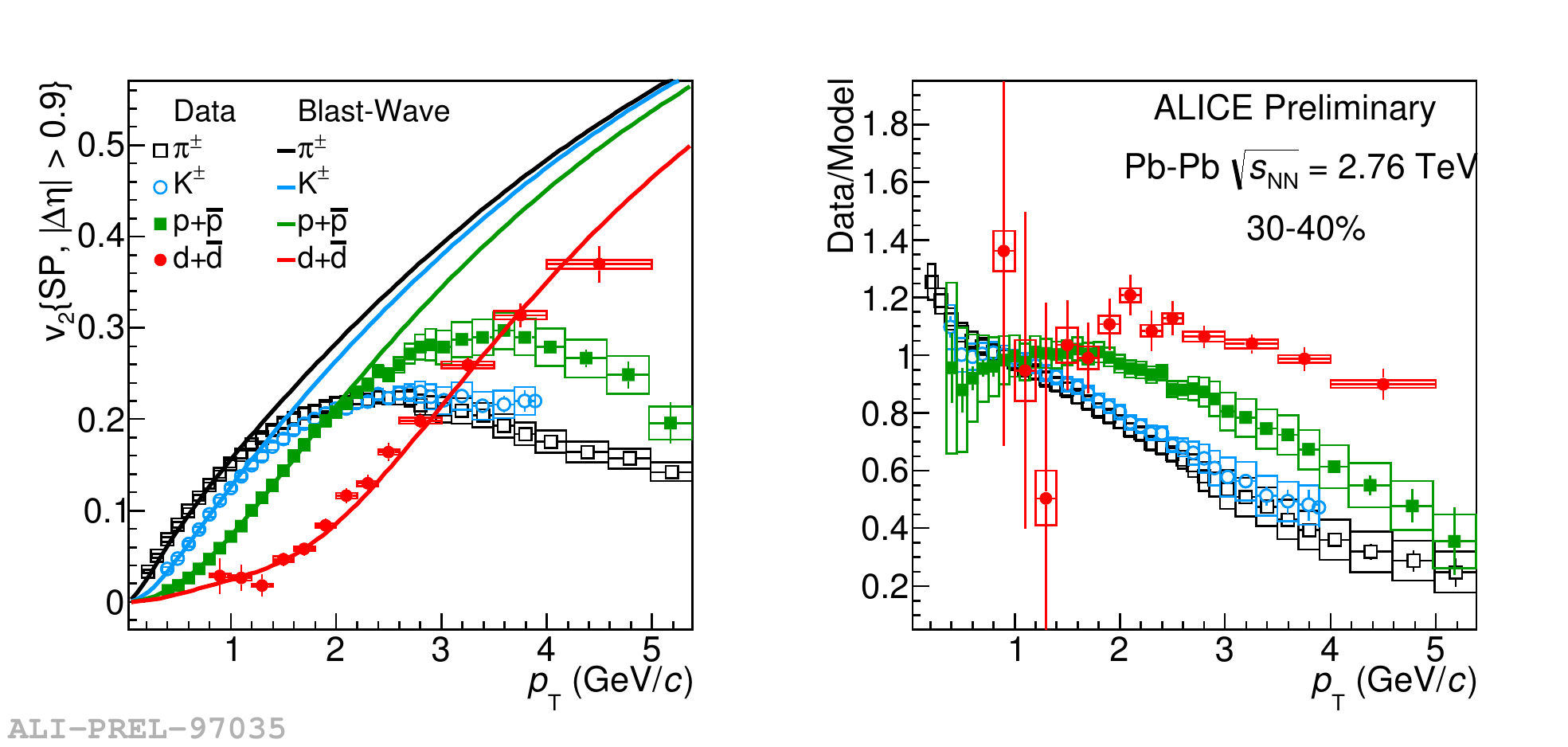}
\caption{Left: Measured \vtwo\ for $\pi^\pm$ (empty squares), K$^\pm$ (empty circles), 
p+$\overline{\rm p}$ (filled squares) and d+\dbar\ (filled circles) with Blast-Wave curves. The parameters 
of the Blast-Wave curve from $\pi^\pm$, K$^\pm$  and p+$\overline{\rm p}$ were used to predict 
deuteron curve. Right: Ratio between data and model: each color corresponds to a 
different particle. Figure copyright CERN, reproduced with permission.}
\label{fig:BWComparison}
\end{center}
\end{figure}

Hydrodynamical calculations are able to reproduce the main features of \vtwo\ for \pt$<$2~\gmom\ 
 \cite{Abelev:2014pua}. 
A simple model based on hydrodynamics is the Blast-Wave model~\cite{Huovinen:2001cy, Adler:2001nb}. 
The measured pions, kaons and protons \pt\ spectra and \vtwo\ (\pt) were fitted, and the parameters 
of the fit were used to predict deuteron \vtwo(\pt). The results for 30-40\% mid-central events are shown in Figure~\ref{fig:BWComparison}. 
For deuterons a good description of the data in the measured \pt\ range and for all 
the measured centralities is observed. 

\section{Conclusions}
\label{sec:conclusions}
The \vtwo\ of deuterons produced in \PbPb\ collisions at \s\ = 2.76 TeV 
was measured up to 5 \gmom\ using the scalar product technique. 
At low \pt\ deuteron \vtwo\ follows mass ordering, indicating a 
more pronounced radial flow in the most central collisions
as observed also for lighter particles. A deviation from A (number of mass) 
and n$_{\rm q}$ (number of constituent quark) scaling at the level of 20\% is observed. 
The Blast-Wave model gives a good description of the measured elliptic flow of deuterons.

\bibliographystyle{elsarticle-num}
\bibliography{<your-bib-database>}

\begin{thebibliography}{00}
\bibitem{Ollitrault:1992bk}
  J.~Y.~Ollitrault, 
  Phys.\ Rev.\ D {\bf 46} (1992) 229, doi:10.1103/PhysRevD.46.229.
\bibitem{Voloshin:1994mz}
  S.~Voloshin and Y.~Zhang,
  Z.\ Phys.\ C {\bf 70} (1996) 665, doi:10.1007/s002880050141 [hep-ph/9407282].
\bibitem{Nonaka:2003ew}
  C.~Nonaka, B.~Muller, M.~Asakawa, S.~A.~Bass and R.~J.~Fries,
  Phys.\ Rev.\ C {\bf 69} (2004) 031902, doi:10.1103/PhysRevC.69.031902 [nucl-th/0312081].
\bibitem{Abelev:2014pua}
  B.~B.~Abelev {\it et al.} [ALICE Collaboration],
  JHEP {\bf 1506} (2015) 190,  doi:10.1007/JHEP06(2015)190 [arXiv:1405.4632 [nucl-ex]].
\bibitem{Huovinen:2001cy}
  P.~Huovinen, P.~F.~Kolb, U.~W.~Heinz, P.~V.~Ruuskanen and S.~A.~Voloshin, 
  Phys.\ Lett.\ B {\bf 503} (2001) 58, doi:10.1016/S0370-2693(01)00219-2  [hep-ph/0101136].
\bibitem{Adler:2001nb}
  C.~Adler {\it et al.} [STAR Collaboration], 
  Phys.\ Rev.\ Lett.\  {\bf 87} (2001) 182301, doi:10.1103/PhysRevLett.87.182301
  [nucl-ex/0107003].
 \bibitem{Abelev:2014ffa}
  B.~B.~Abelev {\it et al.} [ALICE Collaboration],
  Int.\ J.\ Mod.\ Phys.\ A {\bf 29} (2014) 1430044, 
  doi:10.1142/S0217751X14300440
  [arXiv:1402.4476 [nucl-ex]].
\bibitem{Abbas:2013taa}
  E.~Abbas {\it et al.} [ALICE Collaboration],
  JINST {\bf 8} (2013) P10016, 
  doi:10.1088/1748-0221/8/10/P10016
  [arXiv:1306.3130 [nucl-ex]].
\bibitem{Alme:2010ke}
  J.~Alme {\it et al.},
  Nucl.\ Instrum.\ Meth.\ A {\bf 622} (2010) 316, 
  doi:10.1016/j.nima.2010.04.042
  [arXiv:1001.1950 [physics.ins-det]].
\bibitem{Aamodt:2010aa}
  K.~Aamodt {\it et al.} [ALICE Collaboration],
  JINST {\bf 5} (2010) P03003, 
  doi:10.1088/1748-0221/5/03/P03003
  [arXiv:1001.0502 [physics.ins-det]].
\bibitem{Adam:2015vda}
  J.~Adam {\it et al.} [ALICE Collaboration],
  arXiv:1506.08951 [nucl-ex].
\bibitem{Voloshin:2008dg}
  S.~A.~Voloshin, A.~M.~Poskanzer and R.~Snellings,
  in Landolt-Boernstein, Relativistic Heavy Ion Physics, Vol. 1/23, p 5-54 (Springer-Verlag,2010)
  arXiv:0809.2949 [nucl-ex].
\end{thebibliography}

\end{document}